\documentclass[a4paper]{jpconf}
\pdfoutput=1
\usepackage{graphicx}

\usepackage{epsfig}
\usepackage{latexsym}
\usepackage{amsfonts}
\usepackage{amsmath}
\usepackage{amsthm}
\usepackage{amssymb}
\usepackage{amsbsy}
\usepackage{multirow}

\def\calo         {{\cal O}}

\newsavebox{\uuunit}
\sbox{\uuunit}
    {\setlength{\unitlength}{0.825em}
     \begin{picture}(0.6,0.7)
        \thinlines
        \put(0,0){\line(1,0){0.5}}
        \put(0.15,0){\line(0,1){0.7}}
        \put(0.35,0){\line(0,1){0.8}}
       \multiput(0.3,0.8)(-0.04,-0.02){12}{\rule{0.5pt}{0.5pt}}
     \end {picture}}

\def\be{\begin{equation}}
\def\ee{\end{equation}}
\def\bea{\begin{eqnarray}}
\def\eea{\end{eqnarray}}

\newcommand{\beq}{\begin{eqnarray}}
\newcommand{\eeq}{\end{eqnarray}}

\def\L{\Lambda}

\def\f{\phi}

\def\m{\mu}

\def\o{\omega}

\def\p{\pi}
\def\r{\rho}

\def\s{\sigma}

\def\pa{\partial}

\def\to{\rightarrow}

\def\half{{1 \over 2}}
\def\sF{{{ F}\!\!\!\!\hskip.8pt\hbox{\raise1pt\hbox{/}}\,}}
\def\som{{{ \omega}\!\!\!\!\hskip.8pt\hbox{\raise1pt\hbox{/}}\,}}
\def\sJ{{{\rm J}\!\!\!\!\hskip.8pt\hbox{\raise1pt\hbox{/}}\,}}

\begin{document}
\title{Chronology protection in stationary 3D spacetimes}

\author{Joris Raeymaekers}

\address{Institute of Physics of the ASCR\\
Na Slovance 2, 182 21 Prague 8, Czech Republic}

\ead{joris@fzu.cz }

\begin{abstract}
We address chronology protection in stationary, rotationally symmetric spacetimes in 2+1 dimensional gravity,
focusing  on  the case of negative cosmological constant. We show that, if such spacetimes contain closed timelike curves,
they belong to one of the following two classes: (i) bad asymptotics: the boundary itself contains closed timelike curves, or  (ii) unphysical stress tensor in the bulk:
the matter stress tensor violates the null energy condition.
 We illustrate these properties in a class of examples involving rotating dust in anti-de Sitter space and comment on the use of the AdS/CFT correspondence to study chronology protection.
\end{abstract}

\section{Introduction}
The question whether  and how the laws of physics prevent the construction, in principle, of time machines
is a fascinating one
which goes to the heart of our understanding of spacetime geometry and quantum physics.
Hawking's chronology protection conjecture \cite{Hawking:1991nk} states  that the laws of physics prevent the formation
of closed timelike curves (CTCs) that would allow one to travel to one's past (see \cite{Thorne:1992gv,Visser:2002ua} for reviews and further references). At the classical level, evidence
for the conjecture comes from the fact that spacetimes with a compactly generated chronology horizon
require  matter sources that violate the null energy condition (NEC).

Some of the simplest and most well-known examples of spacetimes with closed timelike curves don't have a compactly generated chronology horizon and are therefore
not included in Hawking's original
argument. This is the case for `eternal' time-machines, where the space-time is stationary and every CTC is a member of a one-parameter family of time-translated CTCs.
Examples include the  two earliest known spacetimes with CTCs: the Van Stockum \cite{vanStockum:1937zz} and G\"odel \cite{Godel:1949ga} solutions. Here we will address the issue
of chronology protection in stationary spacetimes in the simplest context: we consider
the case of 2+1 dimensional gravity and consider stationary  spacetimes which are also rotationally symmetric.
For zero cosmological constant, chronology protection in such spacetimes was adressed by Menotti and Seminara \cite{Menotti:1992nz} and was related to violations of the NEC. Here we shall focus instead the case of negative cosmological constant, since we
are ultimately interested in the application of the AdS/CFT correspondence \cite{Maldacena:1997re} to chronology protection. We will comment more on the application
of  AdS/CFT to chronology protection in section \ref{outlook}.

\begin{figure}
\begin{picture}(400,90)
\put(-10,0){\includegraphics[width=200pt]{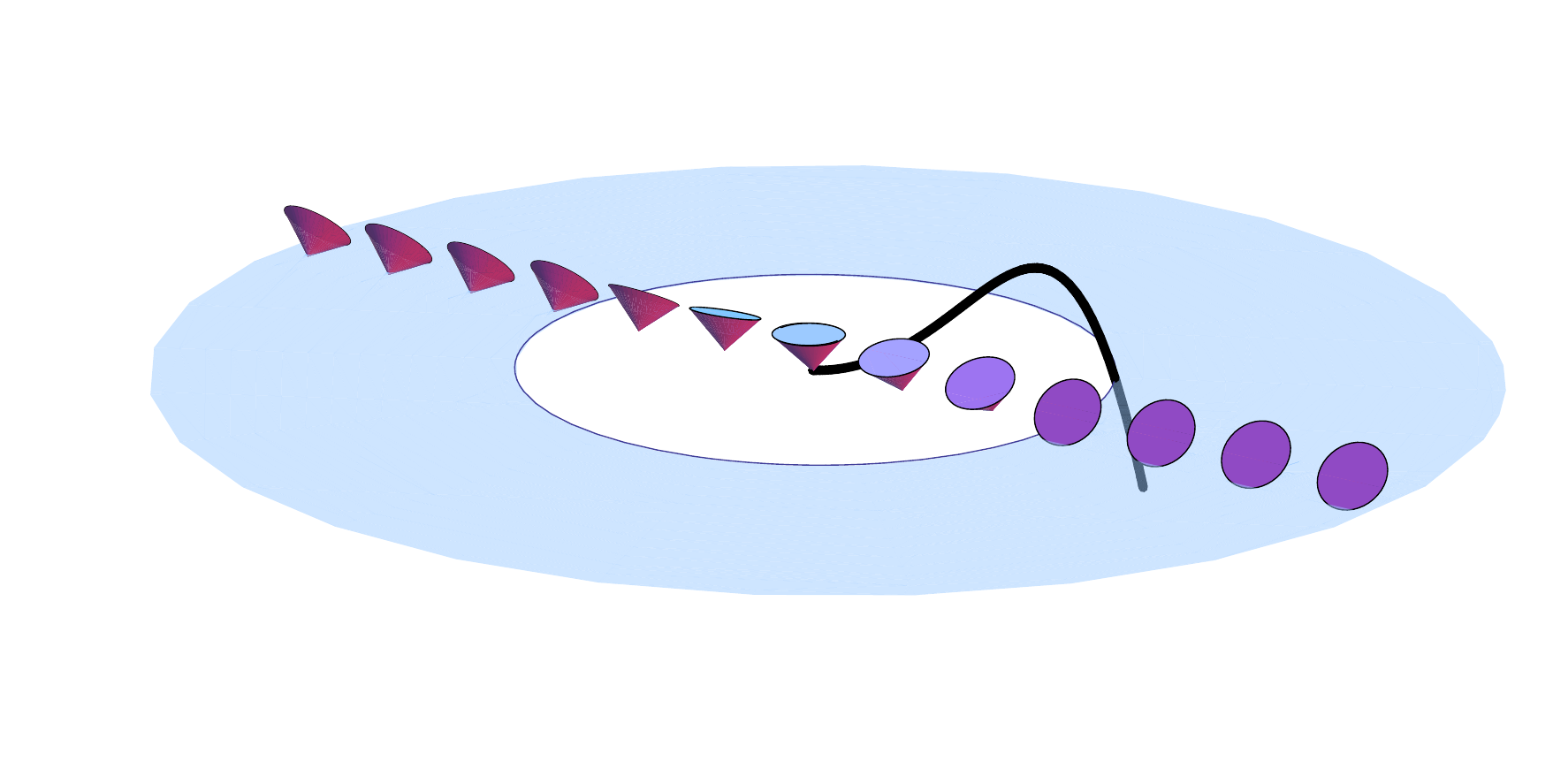}}
\put(180,-5){\includegraphics[width=240pt]{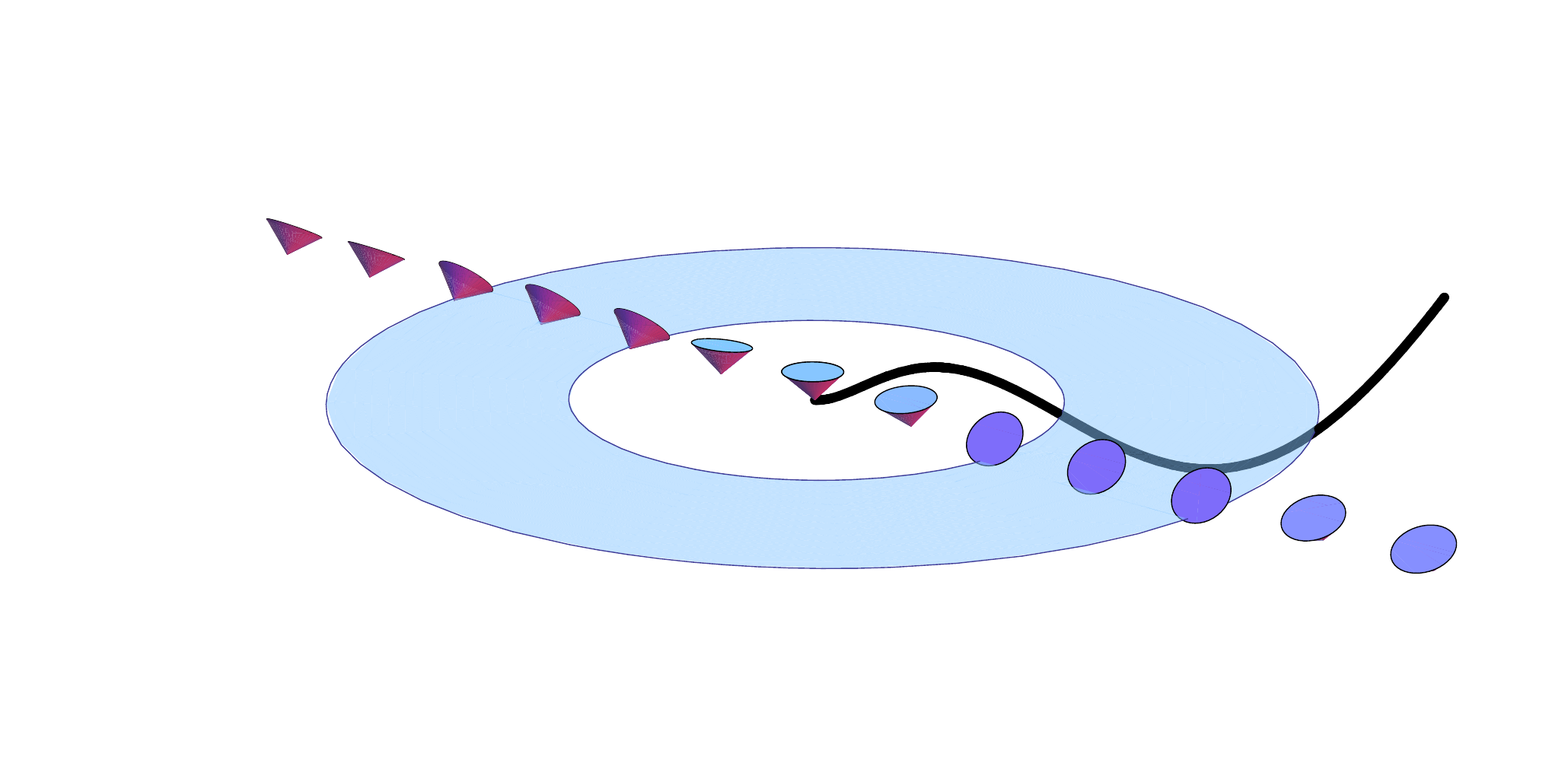}}
\put(90,0){(i)}
\put(290,0){(ii)}
\end{picture}
\caption{The behavior of local lightcones in  a constant time slice in two classes of time machines. The black curve is the is the metric
component $g_{\f\f}$ (see (\ref{metransatz}) below).
}
\label{lcfig}
\end{figure}

Our main result will be to show  that such spacetimes belong to one of the following categories:
\begin{enumerate}
\item the bad asymptotics category: the spacetime has closed timelike curves on the boundary.
\item the unphysical stress tensor  category: the matter stress in the bulk violates the NEC\footnote{Recall that the NEC
requires $n^i T_{ij} n^j \geq 0 $ for every null vector $n^i$.}.
\end{enumerate}
The behavior of
the local lightcones in these two categories of spacetimes  is illustrated in figure \ref{lcfig}.
We will also analyze the implications of violations
of the  NEC  for the conformal weights in the dual CFT, and discuss several examples.
We refer to \cite{Raeymaekers:2011uz} for technical details and a more complete list of references.
\section{Chronology protection for stationary rotationally invariant metrics}\label{chronprotnec}

We consider 2+1-dimensional gravity
in the presence of a cosmological constant. Einstein's  equations are
 \be R_{ij} = 8\p G( T_{ij} - T g_{ij} ) + 2 \L g_{ij}.\label{Einsteq} \ee
We  restrict attention to stationary, rotationally symmetric metrics. We can choose coordinates
$t,\r, \f$ such that the metric takes the form (see e.g. \cite{Deser:1986xf})
\be {ds^2 }
= d\r^2  + g_{tt} dt^2 + 2 g_{t\f} dt d\f + g_{\f\f} d\f^2
.\label{metransatz}
\ee
where $\f$ is identified modulo $2\p$ and $g_{tt}, g_{t\f}, g_{\f\f}$ are
functions of $\r$ that satisfy
\bea
g_{tt} &<& 0\\
- \det g  &=& g_{t\f}^2 - g_{tt}g_{\f\f}   > 0 . \label{posvol}
\eea

We start  by expressing Einstein's equations (\ref{Einsteq}) in a  null triad
$(e^+, e^-, e^\r )$:
\bea
e^\pm &=& {e^{\mp B} \over \sqrt{2}} \left[ \sqrt{-g_{tt} }dt - { g_{t\f} \pm \sqrt{-g} \over \sqrt{-g_{tt}}} d\f\right] \\
e^\r &=& d\r.\label{triad}
\eea
such that the metric takes the form
\be
ds^2 = - 2 e^+ e^- + (e^\r)^2 .
\ee
The function $B(\r ) $ parameterizes the freedom to  locally  Lorentz boost the triad; our argument will be simplified by making a convenient
gauge choice for this function.

The $++$ and $--$  components of Einstein's equations (\ref{Einsteq}) read
\be
R_{\pm\pm} = 8 \p G T_{\pm \pm} \label{einsteqlc}.
\ee
 We now choose the boost function $B$ to satisfy
\be
B ' =  {g_{tt} g_{t\f}' - g_{t\f} g_{tt}'\over 2 g_{tt} \sqrt{-g} }\label{Beq}
\ee
In view of (\ref{posvol}) this differential equation is regular and we are justified in making this choice \footnote{This
gauge choice corresponds to setting to zero the spin connection coefficient $\o_{\r + - }$.}.
The advantage of this gauge is  that the Ricci components $R_{\pm\pm}$ become total divergences
\be
R_{\pm\pm} = \nabla_i v^i_\pm \label{Riccicomps}
\ee
where
\be
v^i_\pm \equiv e_\pm^j \nabla_j e_\pm^i - e_\pm^i \nabla_j e_\pm^j. \label{vdef}
\ee
The $++$ and $--$  components of Einstein's equations (\ref{einsteqlc}) then become simply
\be
\nabla_i v_\pm^i = 8\p G T_{\pm \pm}
\ee
which, since everything depends only on $\r$, can be written as
\be
\pa_\r \left( \sqrt{-g} v_\pm^\r \right) = 8\p G \sqrt{-g} T_{\pm \pm} \label{Komarid}.
\ee
The left hand side is a total derivative while the right-hand side is positive when the NEC holds.
This identity will be our main tool in what follows.
For later use, we record  here also the explicit expression for $\sqrt{-g} v_\pm^\r$:
\be
\sqrt{-g} v_\pm^\r = {e^{\pm 2 B} \over 4 \sqrt{-g}}\left( g_{tt} g_{\f\f}' - g_{\f\f} g_{tt}' +2 {g_{t\f} \mp \sqrt{-g} \over g_{tt} }\left( g_{t\f} g_{t t}' - g_{t t} g_{t \f}'\right)\right).\label{vrho}
\ee

Now we will focus on spacetimes which contain CTCs.
As was shown in \cite{Menotti:1992nz}, this implies that $g_{\f\f}$ must become
negative for some values of $\r$, so that for these values the azimuthal circles are timelike.  We will refer to these timelike $\f$-circles as
azimuthal closed timelike curves (ACTCs).

Now suppose that the spacetime doesn't belong to the category (i) discussed in the Introduction.
This implies that there are no ACTCs for $\r$ sufficiently large, and hence there must be a radius $\r_+$
where $g_{\f\f}$  has a zero and changes sign from negative to positive.
We will further assume  that $g_{\f\f}$ has a second zero at a radius $\r_- < \r_+$ where it changes sign from positive to negative. This
assumption can be be made without much loss of generality for the following reason.
For spacetimes
with a regular axis of symmetry, there must always be such a $\r_-$ since $g_{\f\f}$ is positive in the vicinity of the axis, as one can see
by choosing local inertial coordinates on the axis (see \cite{Mars:1992cm} for a rigorous proof). If there is no axis of symmetry and no
radius  $\r_-$, the metric has a wormhole-like behavior with a second asymptotic region for $\r \to - \infty$ where $g_{\f\f}$ is negative,
and hence would belong to the category (i) contrary to our assumption.

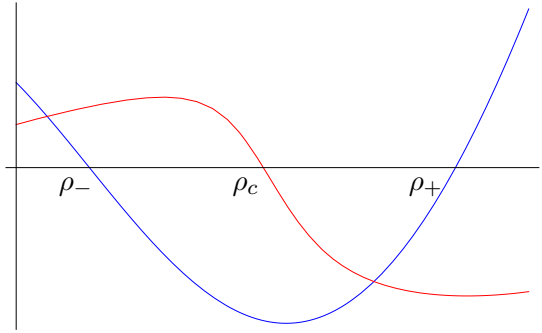
\begin{figure}
\begin{center}
\begin{picture}(300,125)
\put(0,0){\includegraphics[width=200pt]{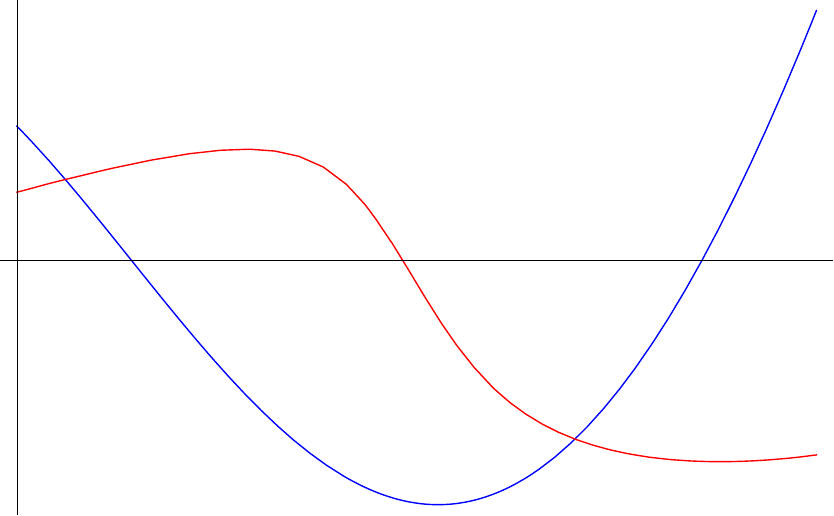}}
\put(20,53){$\r_-$}
\put(85,53){$\r_c$}
\put(151,53){$\r_+$}
\end{picture}
\end{center}
\caption{An example of the behavior of $g_{\f\f}$ (in blue) and $\sqrt{-g} v^\r_\s$ (in red) in a spacetime with localized ACTCs}\label{gffvrhofig}
\end{figure}

Summarized, we have the following behavior in the interval $[ \r_-, \r_+ ]$ (see also Figure \ref{gffvrhofig}):
\bea
g_{\f\f}(\r_-) &=& g_{\f\f}(\r_+) =0\\
g_{\f\f}(\r) &\leq & 0 \qquad {\rm for} \ \r_-\leq\r \leq \r_+\\
g_{\f\f}'(\r_-) &<&0\\
g_{\f\f}'(\r_+) &>&0
\label{rhoplmin}
\eea
This behavior implies some properties of $g_{t\f}$ which we will need later. First of all,  $g_{t\f}$ cannot be zero in $\r_-$ or $\r_+$ because
of (\ref{posvol}). Furthermore,   $g_{t\f}$ has no zeroes in the interval  $[ \r_- , \r_+ ]$, because otherwise the
 metric would not have Minkowski signature.
So we can consistently define
\be
\s \equiv {\rm sign}\, g_{t\f} (\r_- ) = {\rm sign}\, g_{t\f} (\r_+ )\label{sigmadef}
\ee

Now, when we evaluate the quantity $\sqrt{-g} v_\s^\r$ in  $\r_\pm$ using (\ref{vrho}), we see that  only the first term is nonzero:
\be
\sqrt{-g} v_\s^\r (\r_\pm ) = {e^{2 \s B} \over 4 \sqrt{-g}} g_{tt} g_{\f\f}' (\r_\pm).
\ee
Using (\ref{rhoplmin}) we see that $\sqrt{-g} v_\s^\r$ changes sign between $\r_-$ and $\r_+$, so there must be a radius $\r_c$ where it has a zero
(see Figure \ref{gffvrhofig}):
\be
\sqrt{-g} v_\s^\r (\r_c)=0. \label{rhocdef}
\ee
Now we can integrate (\ref{Komarid}) between  $\r_c$ and $\r_+$ in (\ref{rhoplmin}) to get our main identity
\be
8\p G  \int_{\r_c}^{\r_+} d\r \sqrt{-g}  T_{\s\s}  = {e^{2 \s B} \over 4 \sqrt{-g}} g_{tt} g_{\f\f}' (\r_+ )<0 .\label{mainid2}
\ee
We then see that the  null energy condition must be violated on average in the interval $[ \r_c, \r_+ ]$ and the spacetime
belongs to the category (ii) of the Introduction\footnote{Similarly, integrating (\ref{Komarid}) between  $\r_-$ and $\r_c$ one shows that
the NEC is also violated on average in the interval $[ \r_-, \r_c ]$.}.

\section{CTCs and conformal weights}
One expects intuitively that the violation of the NEC in the CTC region represents a negative contribution to the total energy of the system.
In this section we will make this more precise in the case of negative cosmological constant:
we will see that the CTC region represents a negative contribution to either $L_0$ or $\bar L_0$.

First we evaluate the asymptotic values of the quantities $\sqrt{-g} v_\pm^\r$ in the cases where the matter stress tensor vanishes sufficiently fast
for $\r \to \infty$. The metric then asymptotically approaches a vacuum solution
which is stationary and rotationally invariant and is characterize by  two integration constants which can be identified with the mass and
 angular momentum
\cite{Banados:1992wn}. Asymptotically, the metric takes the Fefferman-Graham \cite{FG} form (where $\L = -{1/ L^2}$)
\be
{ds^2 } = d\r^2 +  {  e^{2 \r/L}\over 2 }  \left(-  d   t ^2  + L^2 d\f^2 \right) + { M \over 2} dt^2 - J dt d\f + { M \over 2} L^2 d\f^2 + \calo (e^{- 2 \r/L} )\label{asads}
\ee
The equation (\ref{Beq})  near the boundary implies that $B$ goes to a constant
\be
B = B_\infty + \calo (e^{- 2 \r/L} ).
\ee
Since we have not yet specified the integration constant in (\ref{Beq}), we can fix it by imposing the boundary condition
\be
B_\infty  = \lim_{\r \to \infty} B (\r)  = 0 .\label{bcB}
\ee
For the boundary behavior of $\sqrt{-g} v_\pm^\r$ we then find
\be
\sqrt{-g} v_\pm^\r = {1 \over 2} \left( M \pm {J \over L} \right) + \calo (e^{- 2 \r/L} ).\label{vrhobound}
\ee
Hence the asymptotic  values of $\sqrt{-g} v_+^\r $ and $\sqrt{-g} v_-^\r $ measure essentially the left- and right-moving conformal weights.

We now use this to give a convenient expression for the conformal weight for spaces with  CTCs in the bulk.
Integrating  (\ref{Komarid}) between $\r_c$ and infinity and using (\ref{vrhobound}) we get, e.g. for $\s = +1$:
\be
{ 24\over c }  L_0 = 2 \int_{\r_c}^{\r_+} d\r \sqrt{-g}  T_{++} + 2 \int_{\r_+}^{\infty} d\r \sqrt{-g}  T_{++} + 1   \label{enads}
\ee
For $\s = -1$ we have to replace $L_0$ with $\bar L_0$ and $T_{++}$ with $T_{--}$ in this expression.
The first term on the RHS comes from the CTC region and is negative as we argued already in (\ref{mainid2}). In this precise sense, the CTC
region represents a negative contribution to either $L_0$ (when $\s = 1$) or to $\bar L_0$ (when $\s = -1$).

\section{Examples: rotating dust solutions in AdS}

We will now illustrate the  above properties in some examples. We will find a simple class of analytic solutions involving rotating dust in AdS.
These generalize the three-dimensional G\"odel solution and can be engineered to have  CTCs. In this section we set $\L = - { 1  / 4}$
and consider metrics of the form
\be
{ds^2 } = d\r^2 -d t^2 + 2 l d t d \f + (l'^2 - l^2) d \f^2
\ee
for some function $l(\r)$. Since the determinant of the metric is $|l'|$,  $l'$ cannot vanish (except on the symmetry axis if there is one), and hence
it must have the same sign everywhere.

These metrics satisfy the Einstein equations with a pressureless, rotating dust source:
\be
T^{ab} = R u^a u^b.
\ee
where the energy density $R$ is given in terms of $l$ as
\be
R = 1 - {l''' \over l'}.\label{Req}
\ee
The velocity vector is $u = \pa_t$, so this coordinate system is comoving with the dust.
Within this class of metrics we can construct analytic solutions with the desired behavior of $g_{\f\f}$ by finding a suitable function $l$.

\subsection{The G\"odel universe}
The simplest example of this kind is obtained by taking the energy density of the dust to be constant.
This is achieved by taking
\be
l =  \m \left( 1 - \cosh  { \r \over  \sqrt{\m}}\right).\label{Godell}
\ee
with $\m $ a constant greater than one.  This leads to the energy density $R = {\m -1\over \m}$.

\begin{figure}
\begin{center}
\begin{picture}(120,80)
\put(0,0){\includegraphics[width=120pt]{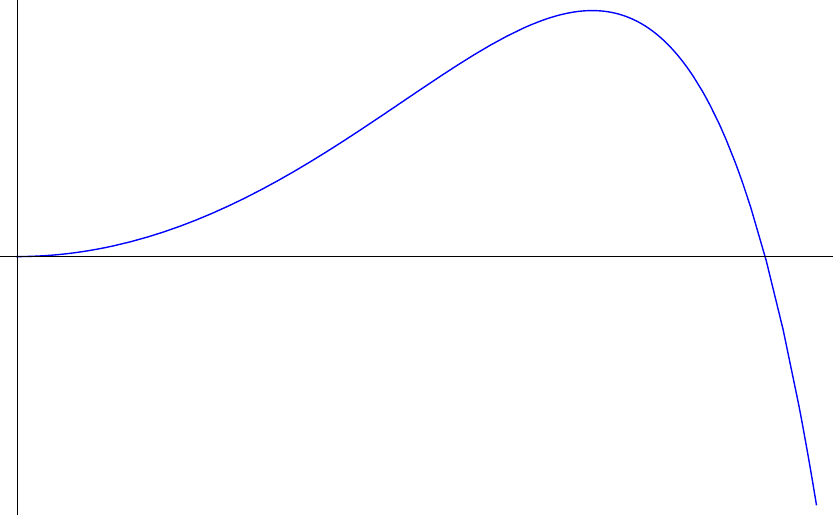}}
\put(73,30){$\r_{max}$}
\put(113,30){$\r_{ctc}$}
\end{picture}
\end{center}
\caption{The metric component $g_{\f\f}$ as a function of $\r$ in the G\"odel universe.}\label{godelgff}
\end{figure}

This solution is known as the 3D G\"odel universe \cite{Rooman:1998xf}, because for the particular value $\m = 2$ it describes the nontrivial
three-dimensional part of G\"odel's original solution \cite{Godel:1949ga}. For $\m, =1$ one recovers the  AdS metric in global coordinates.
The $g_{\f\f}$ component of the metric behaves as follows (see Figure \ref{godelgff}).
  At the radius
\be \r_{max} =\sqrt{\mu } {\rm arccosh}\left(\frac{\mu }{\mu -1}\right) \label{rBdef}\ee it reaches a maximum beyond which it decreases with $\r$.
At the radius \be\r_{CTC} = \sqrt{\m} {\rm arccosh } \left({  \m  + 1\over  \m -1}\right) \label{rctcdef}\ee
$g_{\f\f}$ is zero; beyond this radius it becomes negative and there are ACTCs.
This is an example of the `bad asymptotics' category (i): the NEC is satisfied everywhere but the ACTCs persist to the boundary.

\subsection{A G\"odel dust ball}\label{dustball}
One way of obtaining an asymptotically AdS solution is to take a finite ball of dust with constant density $R$ up to some radius
$\r_0$. The solution for $\r \geq \r_0$ is then a vacuum solution determined by solving the Israel matching conditions \cite{Israel:1966rt} at the edge of the dust ball $\r = \r_0$.
This is the example considered in \cite{Lubo:1998ue,Raeymaekers:2009ij}  which we will here expand on and clarify. When solving the matching problem without
a thin  shell of matter on the edge of the dust ball, one finds that matching onto an asymptotically AdS metric is only possible for radii
$\r_0 \leq \r_{max}$ defined in (\ref{rBdef}). Since we are interested in matchings with $\r_0 \geq \r_{ctc}$ we will include a thin  shell of matter in the matching problem.
In particular, we will consider the matched configuration
\bea
l_- &=&  \m \left( 1 - \cosh  { \r \over  \sqrt{\m}}\right) \qquad {\rm for }\ \r\leq \r_0 \\
l_+ &=& \cosh {\r_0\over \sqrt{\m}} \left( \m - 1 + \cosh (\r - \r_0) \right) - \sqrt{\m} \sinh
{\r_0\over \sqrt{\m}}\sinh (\r - \r_0)-\m \ \  {\rm for }\ \r\geq \r_0 .\label{dustballsing}
\eea
For this configuration, the function $l$ and the metric  are continuous, but  $l'$ and hence also the extrinsic curvature\footnote{In our coordinate system one has $K_{ij} = \half g'_{ij}$.} $K_{ij}$ change sign across the matching surface. From  the Israel matching matching condition  $K^+_{ij} = K^-_{ij} - (T^s_{ij} - T^s g_{ij})$ we deduce that
our configuration corresponds to the singular source
\be
T_{ij}^s = 2 (K_{ij}^- - K^- g_{ij} ).\label{singsource}
\ee

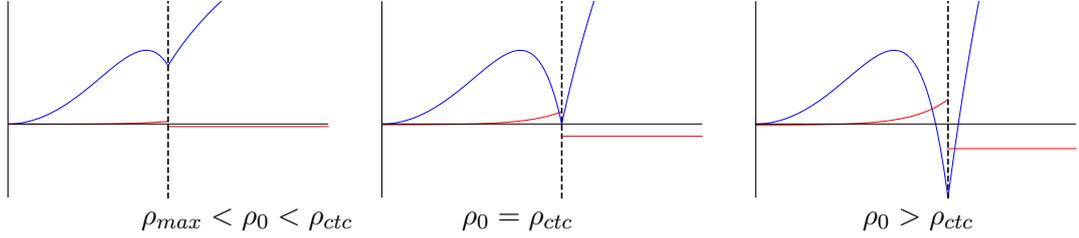
\begin{figure}
\begin{center}
\begin{picture}(370,100)
\put(0,0){\includegraphics[width=120pt]{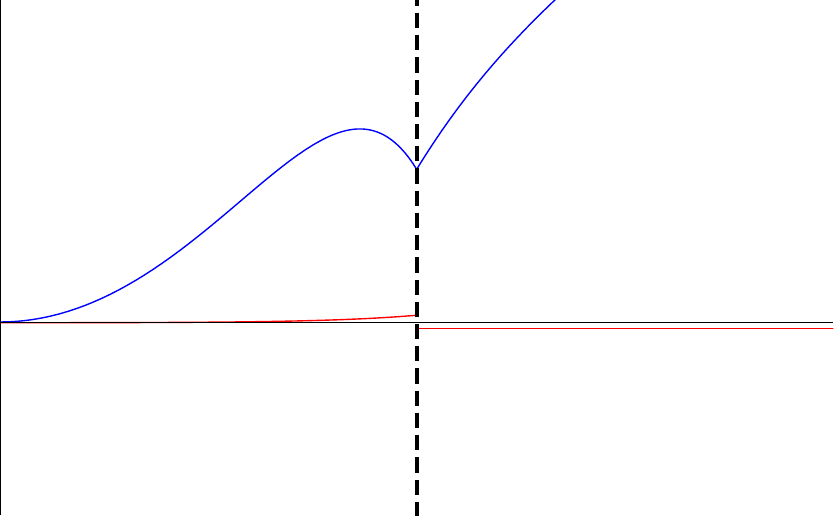}}
\put(140,0){\includegraphics[width=120pt]{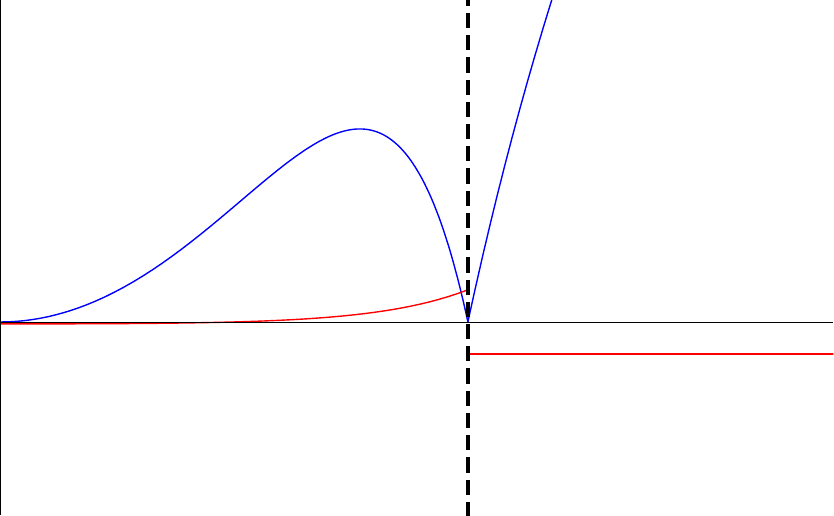}}
\put(280,0){\includegraphics[width=120pt]{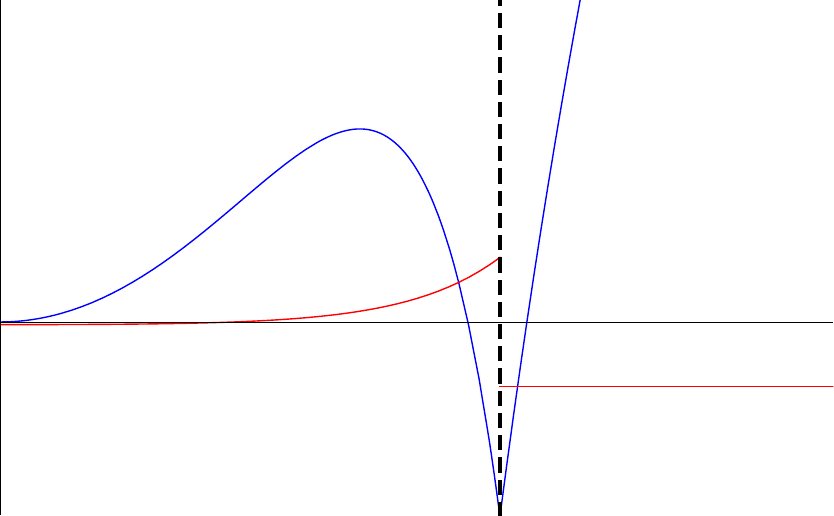}}
\put(50,-10){$\r_{max} <\r_0 < \r_{ctc}$}
\put(170,-10){$\r_0 =\r_{ctc}$}
\put(320,-10){$\r_0 >\r_{ctc}$}
\end{picture}
\end{center}
\caption{The G\"odel dust ball  with a singular source on the edge  for various values of $\r_0$. The blue curve is $g_{\f\f}$, the
red one is $\sqrt{-g} v_+^\r$ and the dotted line denotes the edge $\r_0$ of the dust ball.}\label{singfig}
\end{figure}

The behavior of various quantities in this example are plotted in Figure \ref{singfig}.
For $\r_0 > \r_{ctc}$, the matched configuration has  ACTCs in the bulk\footnote{
In fact,  these were the matched solutions  considered in \cite{Raeymaekers:2009ij} although  the presence of the singular source was overlooked there.}.
 Our argument of section \ref{chronprotnec} tells us that the null energy condition should be violated in the CTC region.
 It is well-known that the stress tensor of the interior G\"odel
space doesn't violate any energy conditions, and we shall presently see that the NEC violation comes from the
 thin shell contribution at the edge. To verify our main identity (\ref{mainid2}), we should put $\r_c = \r_0$ and  include the singular shell contribution to
 the stress tensor.
One then obtains
 \be
 4 \p G \sqrt{-g}  T_{++}^s (\r_0) = {e^{2  B} \over 4 \sqrt{-g}} g_{tt} g_{\f\f}' (\r_+ ) <0.\label{checkid}
\ee
One can check that this equality holds in our example and hence the NEC is violated.

To obtain the total value of $L_0$ we can use our expression  (\ref{enads})
\be
{24 \over c} L_0 =1 + T^s_{++} = 1 - \left(  \m - (\m - 1) \cosh {\r_0\over \sqrt{\m}} \right)^2.
\ee
From (\ref{rctcdef}) we see that the current example has the intriguing property that the total $L_0$ becomes negative precisely when the spacetime contains CTCs, i.e. when $\r_0= \r_{ctc}$. From the point of view of AdS/CFT, such spacetimes are unphysical because unitarity of the dual CFT forbids
negative values of $L_0$.
Since we saw that our example involved a  tuned source on the matching surface, a natural question to ask is whether this behavior is generic.
In the following example we will see that the answer is in the negative: there exist spacetimes with localized ACTCs which have positive conformal weights.

\subsection{A smooth solution with  CTCs in the bulk}\label{smoothexsection}
In this example we display a simple class of smooth solutions which are asymptotically AdS and have localized CTCs.
We take the function $l$ to be of the form
\be
\ln l = \r - {a^2} e^{-\r} + {b - 2 a^4 \over 4} e^{- 2\r} + ( c_1 + c_2 \r + c_3 \r^2)  e^{- 3\r}
\ee
The first three terms guarantee that the metric has asymptotic AdS behavior (\ref{asads})  with the constants $a,b$ related to the total mass and angular momentum.
The coefficients $c_i$ in the fourth term can then be chosen such that $g_{\f\f}$ has two zeroes for small $\r$.

\begin{figure}
\begin{center}
\begin{picture}(200,150)
\includegraphics[height=150pt]{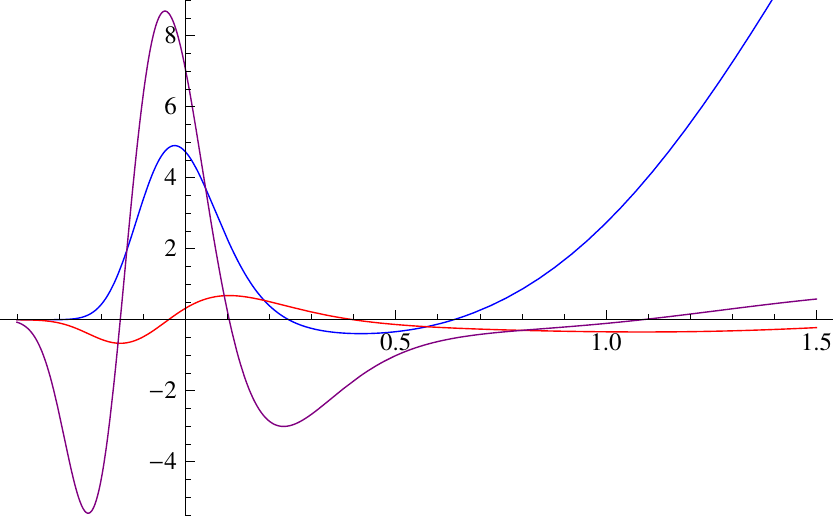}
\end{picture}
\end{center}
\caption{The smooth example with  ACTCs in the bulk discussed in the text. The blue curve is $g_{\f\f}$, the red one is $\sqrt{-g} v^\r_+$ and the purple one is $8\p G \sqrt{-g} T_{++}$ (the latter two are scaled down by a factor 15).}\label{smoothex}
\end{figure}

As an illustrative example, we  take
$a^2 = b  = \half,\  c_1 = - {1 \over 2},\ c_2 =3,\ c_3 = -10$.
One finds that  the corresponding metric has positive conformal weights:
\be
{24 \over c} L_0 = \half \qquad {24 \over c} \bar L_0 = {3\over 4}
\ee
This metric has CTCs between $\r_- \simeq 0.246$ and $\r_+ \simeq 0.638$.
The behavior of various quantities is illustrated in figure \ref{smoothex}.
We see that the NEC is indeed violated in the ACTC region.
In verifying the main identity (\ref{mainid2}) one finds that the radius $\r_c$ defined in (\ref{rhocdef}) is $\r_c \simeq 0.412$, and
\be
\int_{\r_c}^{\r_+} d\r \sqrt{-g}  T_{++}  = {e^{2  B} \over 4 \sqrt{-g}} g_{tt} g_{\f\f}' (\r_+ )\simeq -3.526
 \ee
 We can also verify numerically the expression for $L_0$ in (\ref{enads}): the first term on the RHS, coming  from the ACTC region, contributes approximately -7.052 and the second term,
coming from outside the ACTC region, contributes approximately 6.552, leading indeed to ${24 \over c} L_0= \half$.
We see that in this example, while the ACTC region contributes negatively to $L_0$, the total value of $L_0$ is still positive.

\section{Outlook}\label{outlook}
In this work we showed that simple time machines in asymptotically AdS spaces necessarily violate the NEC, and that the NEC violating region
contributes negatively to the total conformal weight in the dual CFT.
It would  be very interesting to bring the AdS/CFT correspondence to bear on the issue of chronology protection and pinpoint the pathologies
of spacetimes with CTCs.

A first open question is whether CTCs in the bulk are always linked to violation of unitarity in the dual CFT.
In some examples, such as the G\"odel dust ball example of section \ref{dustball}, this is obviously the case:
 the violation of the NEC is so severe that the negative contribution to (\ref{enads}) outweighs the positive one and the total value of $L_0$ is negative.
In other examples, such as the one discussed in section \ref{smoothexsection}, there are CTCs in the bulk while $L_0$ remains positive.
This does not guarantee however that  unitarity is respected.
In such examples, on has to turn on a variety of terms in the metric which are  subleading  at large $\r$
and have the effect of driving $g_{\f\f}$ negative at small values of $\r$ (in our example \ref{smoothexsection}, these are  the terms parameterized by  $c_1, c_2, c_3$).
These subleading terms are expected to encode one-point functions of
 other operators in the CFT than the stress tensor\footnote{For example, in the case of 3D gravity coupled to higher spins \cite{Campoleoni:2010zq}, these subleading terms encode the one-point functions of  primaries of conformal weight greater than two in the dual CFT.}, and hence also to be constrained by unitarity. It would be interesting to explore these constraints further.

Another possible role of AdS/CFT could be to further justify the NEC at the classical level: it is obeyed in most systems but there are exceptions \cite{Barcelo:2002bv}. In the AdS/CFT correspondence, the classical gravity limit describes a certain large $N$ limit of the dual CFT.
 As a consequence, gravity systems that admit CFT duals are constrained already at the classical level by properties of the dual CFT such as unitarity.
It seems promising in this regard that the NEC  has been shown to be related to the c-theorem (and hence ultimately to unitarity) in the dual CFT \cite{Freedman:1999gp}.

And finally, since AdS/CFT allows one  to address quantum corrections in the bulk systematically by taking into account $1/N$ corrections in the boundary theory, it could shed light on the poorly understood issue of  chronology protection in quantum gravity.

\section*{Acknowledgements}

This work was supported  in part by the Czech Science Foundation  grant GACR
P203/11/1388 and in part by the EURYI grant GACR  EYI/07/E010 from EUROHORC and ESF.

\end{document}